\documentclass[twocolumn]{aastex631}
\usepackage{amsmath}
\usepackage{graphicx}

\usepackage{multirow}

\shorttitle{Challenges in forming Phobos and Deimos directly from a splitting of an ancestral single moon}
\shortauthors{Hyodo et al.}
\graphicspath{{./}{figures/}}
\begin{document}

\title{Challenges in forming Phobos and Deimos directly from a splitting of an ancestral single moon}

\correspondingauthor{Ryuki Hyodo}
\email{ryuki.h0525@gmail.com}
\author[0000-0003-4590-0988]{Ryuki Hyodo}
\affiliation{ISAS, JAXA, Sagamihara, Japan}

\author[0000-0001-6702-0872]{Hidenori Genda}
\affiliation{Earth-Life Science Institute, Tokyo Institute of Technology, Meguro-ku, Tokyo 152-8550, Japan}

\author{Ryosuke Sekiguchi}
\affiliation{Earth-Life Science Institute, Tokyo Institute of Technology, Meguro-ku, Tokyo 152-8550, Japan}

\author[0000-0001-5138-230X]{Gustavo Madeira}
\affiliation{Universit\'e de Paris, Institut de Physique du Globe de Paris, CNRS F-75005 Paris, France}
\affiliation{Grupo de Din\^amica Orbital \& Planetologia, S\~ao Paulo State University-UNESP Av. Ariberto Pereira da Cunha, 333, Guaratinguet\'a SP, 12516-410, Brazil}

\author[0000-0002-7442-491X]{S\'ebastien Charnoz}
\affiliation{Universit\'e de Paris, Institut de Physique du Globe de Paris, CNRS F-75005 Paris, France}

\begin{abstract}

The origin and evolution of Martian moons have been intensively debated in recent years. It is proposed that Phobos and Deimos may originate directly from the splitting of an ancestral moon orbiting at around the Martian synchronous orbit. At this hypothetical splitting, the apocenter of the inner moon (presumed as Phobos) and the pericenter of the outer moon (presumed as Deimos) would coincide, in that, their semi-major axes would reside inside and outside the Martian synchronous orbit with non-zero eccentricities, respectively. However, the successive orbital evolution of the two moons is not studied. Here, we perform direct $N$-body orbital integrations of the moons, including the Martian oblateness of the $J_2$ and $J_4$ terms. We show that the two moons, while they precess, likely collide within $\sim 10^4$ years with an impact velocity of $v_{\rm imp} \sim 100-300$ m s$^{-1}$ ($\sim 10-30$ times moons' escape velocity) and with an isotropic impact direction. The impact occurs around the apocenter and the pericenter of the inner and outer moons, respectively, where the timescale of this periodic orbital alignment is regulated by the precession. By performing additional impact simulations, we show that such a high-velocity impact likely results in a disruptive outcome, forming a debris ring at around the Martian synchronous orbit, from which several small moons would accrete. Such an evolutionary path would eventually form a different Martian moon system from the one we see today. Therefore, it seems unlikely that Phobos and Deimos are split directly from a single ancestral moon.

\end{abstract}



\section{Introduction}\label{sec_intro}

The origin of Phobos and Deimos is intensively debated in recent years. Historically, a capture of a passing D-type asteroid, i.e., the capture hypothesis, has been motivated due to spectral similarities to those of the moons \citep{Bur92,Mur91}. Alternatively, a giant impact on Mars could form a debris disk around Mars \citep{Cra11,Hyo17a,Hyo17b,Hyo18a}, i.e., the giant impact hypothesis, from which Phobos and Deimos may accrete as rubble-pile objects \citep{Ros16,Can18}.

It is recently proposed that today's Phobos may not be primordial as a direct consequence of, e.g., either a capture or a giant impact. Instead, after the first generation of Phobos (Phobos's ancestor) is formed, it may have been tidally spiraled inwardly within the Martian Roche limit, recycled into rings via tidal disruption, and then resurrected as a smaller moon via ring's spreading\footnote{In this hypothesis, Deimos is primordial because Deimos orbits outside the Martian synchronous orbit and thus does not tidally spiral inward.}. Today's Phobos may appear after several of this ring-moon recycling evolution \citep{Hes17,Cuk20}, although this view is challenged by the fact that today's Mars does not possess bright particulate rings that are expected to be left behind as a natural consequence of this ring-moon recycling hypothesis (Madeira et al. in prep).

Alternatively, by performing tidal-evolution calculations integrated backward in time, \cite{Bag21} reported that Phobos and Deimos could once have non-zero eccentricities and thus Phobos's apocenter and Deimos's pericenter could cross, while their semi-major axes reside inside and outside the Martian synchronous orbit ($\sim 6 R_{\rm Mars}$ where $R_{\rm Mars}$ is the radius of Mars), respectively. From these findings, they envisioned that Phobos and Deimos were once a single large moon, which was later split into two $-$ as Phobos and Deimos $-$ presumably via a catastrophic impact.

However, their view raises several challenging issues. First, the impact process itself was not studied and thus the likelihood of such an impact, i.e., impact probability and the outcome of impact $-$ whether it splits a single moon into only two with reasonable eccentricity and inclination $-$ were not demonstrated. Second, even if an impact indeed successfully could form two moons as Phobos and Deimos, the successive orbital evolution including mutual interactions (gravity and collision) between the moons were not investigated. The orbital evolution of \cite{Bag21} integrated backward in time was solved based on the orbital elements (e.g., semi-major axis, eccentricity, and inclination) and not on the direct $N$-body approach, neglecting the gravitational interactions and collisions during a moon-moon close encounter. Because Phobos and Deimos initially have orbits that cross each other, the successive orbital evolution may not be as simple as those envisioned and may result in a destructive collision between two moons.

In this study, we especially focus on the second question $-$ the successive orbital evolution after the hypothetical splitting of a single moon into Phobos and Deimos $-$ using a direct $N$-body approach for numerical integration. We focus on the short-term evolution ($<10^4$ years) where the tidal evolution of the moons can be ignored (see Sec.~\ref{sec_tides}). We then show that the two moons in principle collide during the successive orbital evolution within $\sim 10^4$ years. We argue that the impact accompanies a disruptive outcome and the formation of a debris ring. Such an evolutionary path is completely different from the one \cite{Bag21} has envisioned.

This paper is structured as follows. In Section \ref{sec_method}, we describe our methods of orbital integration. In Section \ref{sec_results}, we present the numerical results of the orbital integrations of the two moons that are hypothetically split from a single ancestral moon and show that the two moons likely collide. In Section \ref{sec_impact}, we perform additional impact simulations of the two moons and present that the outcome is disruptive, forming a debris ring. In Section \ref{sec_discussion}, we discuss the dynamical fate of the debris ring and envision the formation of multiple moons (more than three moons). Finally, Section \ref{sec_summary} summarizes our conclusions.

\section{Numerical method}\label{sec_method}

\begin{deluxetable*}{ccc}
\tablenum{1}
\tablecaption{Parameters used in this study \label{table}}
\tablewidth{0pt}
\tablehead{
\colhead{name} & \colhead{mass [kg]} & \colhead{mean radius [km]} 
}
\startdata
Mars     &  $6.39 \times 10^{23}$  & 3389.5  \\
Phobos &  $1.06 \times 10^{16}$  & 11.3      \\
Deimos &  $1.48 \times 10^{15}$  & 6.3        \\
\enddata
\end{deluxetable*}

\begin{figure*}[t!]
\plotone{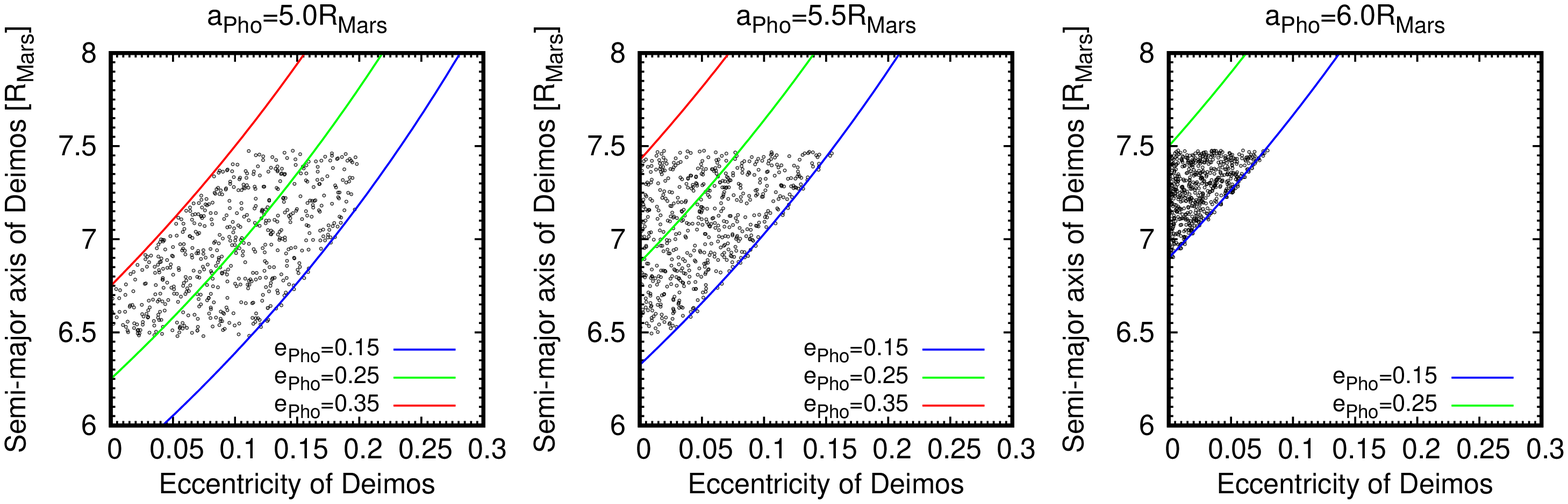}
\caption{Initial distribution of $e_{\rm Dei}$ and $a_{\rm Dei}$. Each black point is the initial conditions of our numerical simulations (600 points). Left, middle, and right panels show cases of $a_{\rm Pho}=5.0$, $5.5$, and $6.0R_{\rm Mars}$, respectively. We set $e_{\rm Pho}=0.15-0.35$, $e_{\rm Dei}=0.0-0.2$, and $a_{\rm Dei}=6.5-7.5R_{\rm Mars}$, following \cite{Bag21}. $a_{\rm Dei}$ is obtained from Eq.~(\ref{eq_ini}) with $e_{\rm pho}$ and $e_{\rm Dei}$ randomly distributed within the ranges. The blue, green, and red curves indicate $a_{\rm Dei}$ for $e_{\rm Pho}=0.15$, $0.25$, and $0.35$, respectively. \label{fig_initial}}
\end{figure*}

\subsection{Orbital calculation}
We performed three-body (Mars-Phobos-Deimos) numerical simulations. Orbits of the bodies were integrated by using the fourth-order Hermite method \citep{Mak92,Kok04} and the numerical code was originally developed in previous studies \citep{Hyo16}. We included the second-order and fourth-order oblateness moments of Mars (i.e., $J_2$ and $J_4$)\footnote{The $J_3$ term could periodically change eccentricity and inclination but it is negligible for our chosen parameters \citep{Liu21}.}. The equation of motions in this study ($xy$-plane is the Martian equatorial plane) are 
\begin{align}
	\ddot{x}_i &= -GM_{\rm Mars} \displaystyle \frac{x_i}{|r_i|^3} \left( 1 - J_2\Psi_{i2} -J_4 \Psi_{i4} \right) - \sum_{j \neq i} G m_j  \displaystyle \frac{x_i - x_j}{r_{ij}^3}  \\	
	\ddot{y}_i &= -GM_{\rm Mars} \displaystyle \frac{y_i}{|r_i|^3} \left( 1 - J_2\Psi_{i2} -J_4 \Psi_{i4} \right) - \sum_{j \neq i} G m_j  \displaystyle \frac{y_i - y_j}{r_{ij}^3} 
\end{align}
\begin{align}
	\ddot{z}_i = -GM_{\rm Mars}  \displaystyle \frac{z_i}{|r_i|^3} \left( 1 - J_2\Psi_{i2} -J_4 \Psi_{i4} + J_2\Phi_{i2} + J_4\Phi_{i4} \right) \nonumber \\ 
	- \sum_{j \neq i} G m_j  \displaystyle \frac{z_i - z_j}{r_{ij}^3} ,
 \end{align}
where $G$ and $M_{\rm Mars}$ are the gravitational constant and the mass of Mars, respectively. Subscripts of $i$ and $j$ indicate Phobos or Deimos. $r_{i}=\left( x_{i}, y_{i}, z_{i} \right)$ is the position vector and $r_{\rm ij}=|r_{\rm i} - r_{\rm j}|$. $m_{\rm j}$ is the mass of Phobos or Deimos. $\Psi_{i2}$, $\Psi_{i4}$,  $\Phi_{i2}$, and $\Phi_{i4}$ are \citep[][]{Sin85} 
\begin{eqnarray}
	\Psi_{i2} &=& \displaystyle \frac{R_{\rm Mars}^2}{r_i^2} P'_3 \left( \displaystyle \frac{z_i}{r_i} \right) \\
	\Psi_{i4} &=& \displaystyle \frac{R_{\rm Mars}^4}{r_i^4} P'_5 \left( \displaystyle \frac{z_i}{r_i} \right) \\
	\Phi_{i2} &=& 3\displaystyle \frac{R_{\rm Mars}^2}{r_i^2}, \hspace{2em} \\
	\Phi_{i4} &=& \displaystyle \frac{R_{\rm Mars}^4}{r_i^4} Q_4 \left(\displaystyle \frac{z_i}{r_i} \right) ,
\end{eqnarray} 
where the $P_{\rm n}'(x)$ terms are the derivative of the Legendre polynomial, $P_{n}(x)$, and $P'_4(x)=xQ_4(x)$ given as
\begin{eqnarray}
	P'_3(x) &=& \displaystyle \frac{15}{2}x^2 - \displaystyle \frac{3}{2} \\
	P'_5(x) &=& \displaystyle \frac{315}{8}x^4  - \displaystyle \frac{105}{4}x^2 + \displaystyle \frac{15}{8} \\
	Q_4(x) &=& \displaystyle \frac{35}{2}x^2 - \displaystyle \frac{15}{2} .
\end{eqnarray} 
Here, $J_{2}= 1.96 \times 10^{-3}$ and $J_{4} = -1.54 \times 10^{-5}$ \citep{Yor95,Liu11}.\\

We note that the other external perturbation forces may slightly change the eccentricities of the moons. The most important perturbation could be evection \citep[][in analogy to Triton]{Gol89}. The amplitude of the periodic change in $e$ due to evection is of the order of $\sim (n_{\rm p}/n_{\rm s}) e$, where $n_{\rm p}$ and $n_{\rm s}$ are the mean motions of the planet and the satellite, respectively \citep[][see their Eq.~(19)]{Cuk04}. As $n_{\rm p}/n_{\rm s} \lesssim 10^{-3}$ for the cases of Phobos and Deimos, the amplitude of change of pericenter of the moons is estimated to be smaller than the size of the moons, making our calculations largely unaffected. Thus, we neglected evection in this study.

\subsection{Initial conditions} \label{sec_initial}
Following the view of \cite{Bag21} in that Phobos and Deimos are split from a single progenitor moon, we set Phobos's apocenter and Deimos's pericenter initially equal as 
\begin{equation}
	r_{\rm Pho,apo} \left( = a_{\rm Pho} \left( 1 + e_{\rm Pho} \right) \right) = r_{\rm Dei,peri} \left( = a_{\rm Dei} \left( 1 - e_{\rm Dei} \right) \right) ,
\label{eq_ini}
\end{equation}
where $a$ and $e$ are semi-major axis and eccentricity, respectively. In this paper, the subscripts of "Pho" and "Dei" indicate Phobos and Deimos, respectively. For the orbits of Phobos and Deimos to initially be in touch, we set the argument of periapsis, $\omega$, and the longitude of ascending node, $\Omega$, as $|\omega_{\rm Pho} - \omega_{\rm Dei}| = \pi$ and $\Omega_{\rm Pho}=\Omega_{\rm Dei}$, respectively. Inclinations, $i$, of Phobos and Deimos hardly change in billions of years of tidal evolution. We used $i_{\rm Pho}=0.021$ rad ($\sim 1.2$ deg) and $i_{\rm Dei}=0.015$ rad ($\sim 0.86$ deg) that would be the largest difference between those of Phobos and Deimos \citep[see][]{Bag21}. A smaller difference in their inclinations indicates that the two orbital planes are more coincident, leading to a more frequent close encounter. Even if we use today's values (i.e., the Laplace plane of Deimos is not the same as that of Phobos or the Martian equator), it should not significantly affect the collisional timescale reported in this study. This is because the inclination of Deimos to the Laplace plane ($\sim 2$ deg) is larger than the Laplace plane tilt ($< 1$ deg).

We initially randomized the eccentric anomaly that defines the position of a body along a given elliptic Kepler orbit. Table \ref{table} lists other physical parameters used in this study. We note, importantly, that the two orbits initially could be those ``crossed" (i.e., $r_{\rm Pho,apo} > r_{\rm Dei,peri}$) at the hypothetical splitting, although here they were set to ``touch" each other (i.e., $r_{\rm Pho,apo} = r_{\rm Dei,peri}$; Eq.~(\ref{eq_ini})). Such initial orbits would be more prone to collide as the orbits cross.

\cite{Bag21} reported that, at the hypothetical splitting of a single large moon into Phobos and Deimos (i.e., the initial condition of our orbital integrations), $a_{\rm Pho} \sim 5-6 R_{\rm Mars}$, $e_{\rm Pho} \sim 0.15-0.35$, and $e_{\rm Dei} \sim 0.0-0.2$. The minimum and maximum $a_{\rm Dei}$ are $\sim 6.5R_{\rm Mars}$ and $7.5R_{\rm Mars}$, respectively \citep{Bag21}. 

We fixed $a_{\rm Pho}=5.0$, $5.5$, and $6R_{\rm Mars}$ and randomly distributed $e_{\rm Pho}$ and $e_{\rm Dei}$ within the aforementioned ranges to create the initial conditions for our numerical simulations. Using these values, $a_{\rm Dei}$ is derived from Eq.~(\ref{eq_ini}). Figure \ref{fig_initial} shows the initial conditions of our numerical simulations (black points). We performed 600 simulations each for $a_{\rm Pho}=5.0$, $5.5$, and $6.0R_{\rm Mars}$. We terminated the simulations when a collision of two moons is detected or when simulation time exceeds $1 \times 10^4$ years.

\section{Results}\label{sec_results}

\subsection{General outcome after splitting} \label{sec_general}

\begin{figure*}[t!]
\centering
\includegraphics[width=1.5\columnwidth]{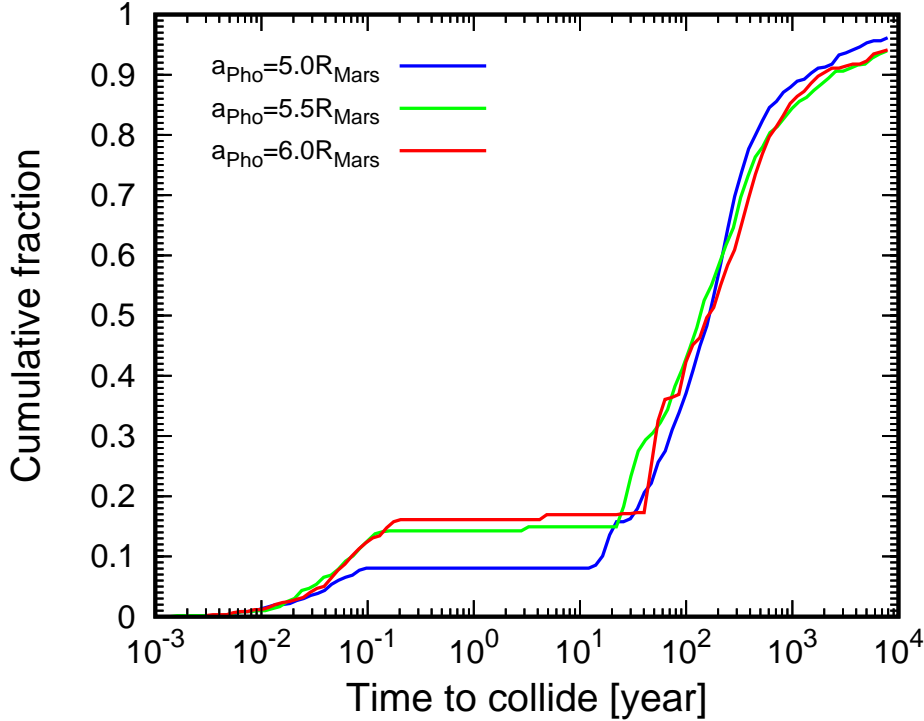}
\caption{Cumulative distribution of the time taken to collide in years. Blue, green, and red lines represent cases of $a_{\rm Pho}=5.0$, $5.5$, and $6.0R_{\rm Mars}$, respectively. Two distinct timescales of collisions are seen; $t_{\rm col} \sim 10^{-2} - 10^{-1}$ years and $t_{\rm col} \gtrsim 30$ years. Less than $10$ \% of our runs  at $1 \times 10^4$ years  still do not experience a collision between the moons. \label{fig_time_to_collide}}
\end{figure*}

In short, after the hypothetical splitting of a single moon into two, presumably as Phobos and Deimos, these two moons most likely collide at around the apocenter of Phobos and at the pericenter of Deimos. More than $>90$\% of our simulations result in a collision (no specific correlation exists between the outcome and the initial conditions as three-body problem has chaotic behaviour). Figure \ref{fig_time_to_collide} shows the cumulative distribution of the time of collision between the moons, $t_{\rm col}$, since the start of our numerical simulations for different initial semi-major axes ($a_{\rm Pho} =5.0$, $5.5$, and $6.0 R_{\rm Mars})$. Two distinct cases are observed: $t_{\rm col} \sim 10^{-2} - 10^{-1}$ years ($\sim 10-20$\% of runs) and $t_{\rm col} \gtrsim 30$ years ($\sim 70-80$\% of runs).

Cases of $t_{\rm col} \sim 10^{-2} - 10^{-1}$ years can be explained by the following two timescales. First, because Phobos's apocenter and Deimos's pericenter are initially in touch (Eq.~(\ref{eq_ini})) but they have different semi-major axes, they can potentially collide with a timescale of their synodic period. The synodic period is given as 
\begin{eqnarray}
	T_{\rm syn} &=& \frac{2\pi a}{\frac{3}{2}\Delta a \Omega_{\rm K}} \nonumber \\
		           &\sim& 0.01 {\, \rm years} \left( \frac{a}{6R_{\rm Mars}} \right)^{5/2}\left( \frac{\Delta a}{R_{\rm Mars}} \right)^{-1} ,
\label{eq_tsyno}
\end{eqnarray}
where $\Delta a$ is the difference in semi-major axes and $\Omega_{\rm K}$ is the Keplerian orbital frequency. For typical values of $a \sim 6R_{\rm Mars}$ and $\Delta a \sim 1R_{\rm Mars}$, $t_{\rm syn} \sim 0.01$ years. 

Second, when precession takes place, the argument of pericenter, $\omega$, and  the longitude of ascending node, $\Omega$, of Phobos and Deimos relatively change. This leads to a misalignment of the pericenter-to-apocenter from the initial configuration. These precession rates, $\dot{\omega}$ and $\dot{\Omega}$, are dominated by the $J_2$ term (because $J_2 \gg J_4$) and are described as \citep{Kau66,Dan92}
\begin{eqnarray}
	\dot{\omega} &=& \frac{3 n}{\left( 1-e^2 \right)^2} \left( \frac{R_{\rm Mars}}{a} \right)^2 \left( 1-\frac{5}{4}\sin^2(i) \right) J_2 \label{eq_pre_omega} \\
	\dot{\Omega} &=&  -\frac{3n \cos(i)}{2\left( 1-e^2 \right)^2} \left( \frac{R_{\rm Mars}}{a} \right)^2 J_2,
\label{eq_pre_Omega}
\end{eqnarray}
where $n=\sqrt{GM_{\rm Mars}/a^3}$ is the orbital mean motion. For a small eccentricity and inclination, the synodic periods of the relative precession timescale of the argument of pericenter, $T_{\rm syn,\omega}$, and of the longitude of ascending node, $T_{\rm syn,\Omega}$, between two moons can be written as  
\begin{align}
 	T_{\rm syn,\omega} &\equiv \frac{2\pi}{ \frac{d \dot{\omega}}{da} \Delta a} \sim 29.4 {\, \rm years}\left( \frac{a}{6R_{\rm Mars}} \right)^{9/2} \left( \frac{\Delta a}{R_{\rm Mars}} \right)^{-1} \label{eq_time_pre_omega}\\
 	T_{\rm syn,\Omega} &\equiv \frac{2\pi}{ \frac{d \dot{\Omega}}{da} \Delta a} \sim 58.8 {\, \rm years} \left( \frac{a}{6R_{\rm Mars}} \right)^{9/2} \left( \frac{\Delta a}{R_{\rm Mars}} \right)^{-1} .
\label{eq_time_pre_Omega}
\end{align}
As $\omega$ precesses faster than $\Omega$ (Eqs.~(\ref{eq_pre_omega})-(\ref{eq_pre_Omega}) and Eqs.~(\ref{eq_time_pre_omega})-(\ref{eq_time_pre_Omega})), the relative precession of $\omega$ initially dominates a misalignment of the pericenter-to-apocenter. 

When the change in the relative radial distance (i.e., the difference in the radial distances between Phobos and Deimos at the synodic period) through the relative precession of $\omega$ becomes larger than the sum of the moons' radii, $R_{\rm moon}$, the orbits of the two moons are no longer in touch and a collision does not anymore occur. During their relative precession, the minimum distance between Phobos and Deimos at the synodic period changes from $0$ (i.e., the initial pericenter-to-apocenter alignment) to $a_{\rm Dei}\left( 1+ e_{\rm Dei} \right) - a_{\rm Pho}\left( 1+ e_{\rm Pho} \right) \sim a_{\rm Dei} - a_{\rm Pho}$ for nearly circular orbits (i.e., when they relatively precess by $\pi$ from the initial configuration). 

Thus, assuming a steady change, the critical time, $T_{\rm sep,ini}$, needed to radially separates the two moons from the initial configuration of the pericenter-to-apocenter alignment via the relative precession of $\omega$ is given as
\begin{align}
	T_{\rm sep,ini} \sim \frac{T_{\rm syn,\omega}}{2 \pi} \theta_{\rm cri} \sim \frac{T_{\rm syn,\omega}}{2 \pi} \frac{R_{\rm moon} \pi}{a_{\rm Dei} - a_{\rm Pho}} \sim 0.1  {\, \rm years} ,
\end{align}
where $\theta_{\rm cri}$ is the critical angle between the arguments of periapsis of Phobos and Deimos (in radian) to physically separate the two moons ($R_{\rm moon} \sim \left( a_{\rm Dei} - a_{\rm Pho} \right) \frac{\theta_{\rm cri}}{\pi}$). Here, $R_{\rm moon} \sim 20$ km and $a_{\rm Dei} - a_{\rm Pho} \sim R_{\rm Mars}$ are used. Hence, $t_{\rm col} \sim 10^{-2} - 10^{-1}$ years indicates that Phobos and Deimos collide just after the start of numerical simulations before the Martian oblateness (mainly by $J_{2}$) precesses their orbits large enough to radially separate them.

Cases of $t_{\rm col} \gtrsim 30$ years can be explained as follows. When Phobos and Deimos avoid a collision during the first $\sim 10^{-2} - 10^{-1}$ years, an orbital precession due to Martian oblateness (mainly by $J_{2}$) effectively changes the moon's relative orbital configurations so that their orbits no longer cross (after $\sim 10^{-1}$ years). Precession of $\omega$ changes the direction of the pericenter, while that of $\Omega$ changes the position where the orbits of the moons pass through the reference plane. 

Thus, assuming no significant change in orbits occur during close encounters, the orbits of the two moons do not cross again until (1) $\Omega_{\rm Pho}=\Omega_{\rm Dei}$ via the relative precession of $\Omega$ and (2) the apocenter of Phobos is pointed towards the pericenter of Deimos, i.e., $|\omega_{\rm Pho}-\omega_{\rm Dei}|=\pi$, via the relative precession of $\omega$. 

The synodic periods of the relative precession timescale of $\omega$ and $\Omega$ are given in Eqs.~(\ref{eq_time_pre_omega}) and (\ref{eq_time_pre_Omega}). These two timescales indicate that Phobos and Deimos have a chance to collide every $\sim 30$ years, which is consistently observed in the results of numerical simulations (i.e., $t_{\rm col} \gtrsim 30$ years).

\subsection{Impact conditions} \label{sec_condition}

\begin{figure*}[t!]
\plotone{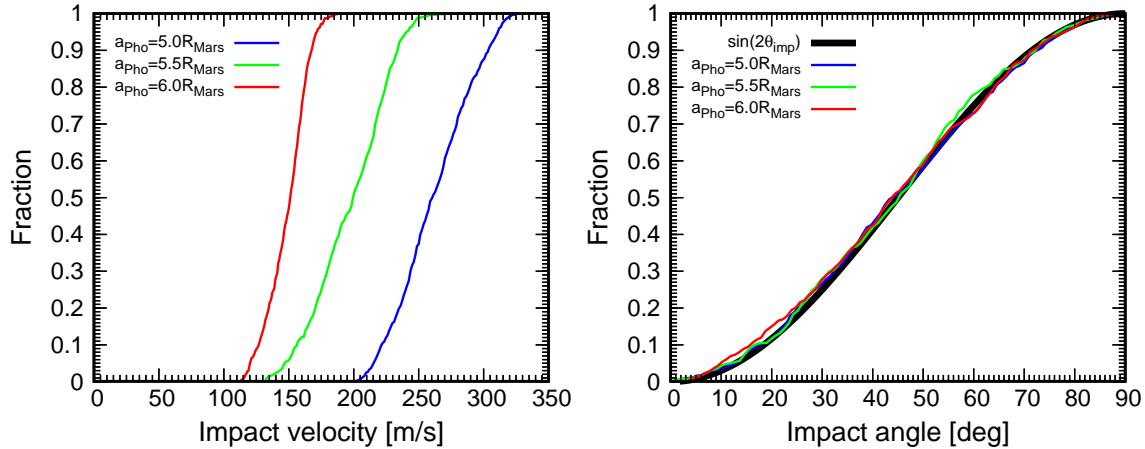}
\caption{Cumulative distributions of impact velocity (left) and impact angle (right) obtained from our numerical simulations. Blue, green, and red colors indicate $a_{\rm Pho}=5.0$, $5.5$, and $6.0R_{\rm Mars}$, respectively. In the right panel, the black curve shows the cumulative of $P(\theta_{\rm imp})=\sin(2\theta_{\rm imp})$ which has a peak at $\theta_{\rm imp} = 45$ deg.} \label{fig_impact_conditions}
\end{figure*}

In Sec.~\ref{sec_general}, most of our numerical simulations (more than $90$\% of our runs) showed that the two moons that are split from a single moon envisioned by \cite{Bag21} eventually collide with each other within $\sim 10^4$ years. Here, by further analyzing the data of our numerical simulations, we show the impact conditions at the collisions (i.e., impact velocity, $v_{\rm imp}$, and impact angle, $\theta_{\rm imp}$).

Figure \ref{fig_impact_conditions} shows the cumulative distributions of the impact velocity (left) and the impact angle (right). Blue, green, and red colors indicate cases of $a_{\rm Pho}=5.0$, $5.5$, and $6.0R_{\rm Mars}$ ($a_{\rm Dei} > 6.5R_{\rm Mars}$; see Sec.~\ref{sec_initial}), respectively. 

As $a_{\rm Pho}$ becomes smaller, the impact velocity becomes larger. This is because the Keplerian velocity depends on $a^{-1/2}$ and because the relative velocity between Phobos and Deimos increases with increasing the difference in their semi-major axes (see also Fig.~\ref{fig_initial}). For $a_{\rm Pho}=5-6R_{\rm Mars}$ and $a_{\rm Dei}=6.5-7.5R_{\rm Mars}$, $v_{\rm imp} \simeq 100-300$ m s$^{-1}$. This is reasonably understood by considering the random velocity, $v_{\rm ran}$, as $v_{\rm ran} \simeq \sqrt{e^2 + i^2} v_{\rm K} \sim 100-400$ m s$^{-1}$ for typical values of the Keplerian velocity of $v_{\rm K} \simeq 1450$ m s$^{-1}$ at $a=6R_{\rm Mars}$ and of $e \sim 0.1-0.3$ with $i \sim 0$. 

The distribution of the impact angle, defined to be $\theta_{\rm imp} = 0$ deg for a head-on collision and $\theta_{\rm imp} = 90$ deg for a perfect grazing impact, indicates that its probability distribution follows nearly $P(\theta_{\rm imp})=\sin(2\theta_{\rm imp})$ with a peak at $\theta_{\rm imp} = 45$ deg (the black line in the right panel of Figure \ref{fig_impact_conditions}). Thus, the impact direction is nearly an isotropic distribution.

\begin{figure*}[t!]
\plotone{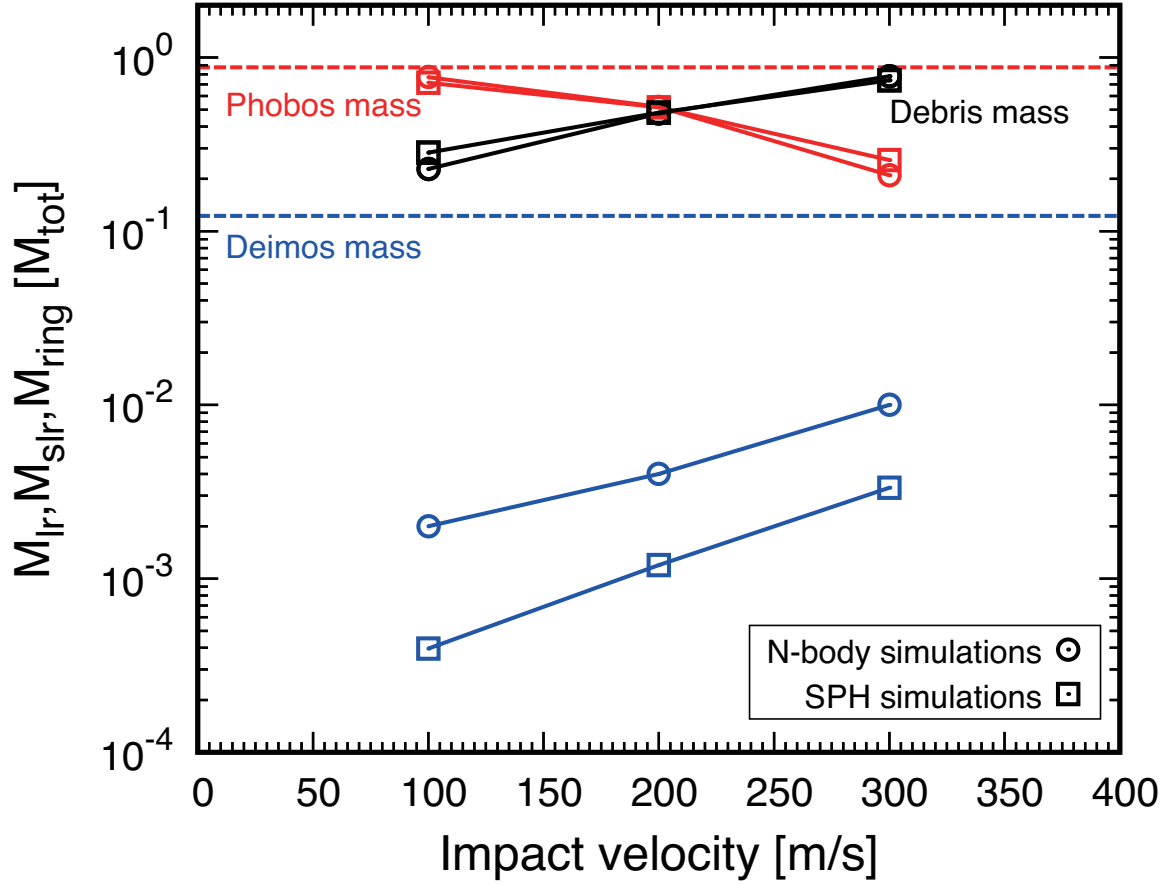}
\caption{Outcomes of collision between Phobos and Deimos. Masses of the largest remnant ($M_{\rm lr}$; red points), the second largest remnant ($M_{\rm slr}$; blue points), and the debris ring ($M_{\rm ring} = M_{\rm tot} - M_{\rm lr} - M_{\rm slr}$ where $M_{\rm tot}=m_{\rm Pho}+m_{\rm Dei}$; black points) as a function of impact velocity are shown. The red and blue horizontal dashed lines indicate masses of Phobos and Deimos, respectively. The open circles are the results of $N$-body simulations \citep[][the cases of the impactor-to-target mass ratio of $\gamma=0.1$ and $\theta_{\rm imp}=45$ deg]{Lei12}. The open squares are the results obtained from our SPH impact simulations. In our impact simulations, the masses of the target and the impactor are $m_{\rm Pho}$ and $m_{\rm Dei}$, respectively. $\theta_{\rm imp}=45$ deg is used. \label{fig_impact}}
\end{figure*}

\section{Fate of impact between two moons} \label{sec_impact}
In Section \ref{sec_results}, we show that the hypothetical two moons, presumably as Phobos and Deimos, that are split from a single ancestral moon collide during the successive orbital evolution. Collision velocity is $v_{\rm imp} \sim 100-300$ m s$^{-1}$, that is, $v_{\rm imp}\sim 10-30 v_{\rm esc}$ (the escape velocities of Phobos and Deimos are $v_{\rm esc} \sim 5-10$ m s$^{-1}$). Such a high-velocity collision may result in a disruptive outcome, while their small mass ratio between the two moons (their mass ratio is $\gamma \simeq 0.1$) may lead to a less catastrophic outcome for the larger one (i.e., target) compared to the case of an impact between comparable masses \citep{Lei12}.

Here, we additionally performed 3D impact simulations, using the smoothed particle hydrodynamics (SPH) approach \citep{Mon92}, to examine the typical outcome of an impact. We employed the impact velocity $v_{\rm imp}=100-300$ m s$^{-1}$ and the impact angle $\theta_{\rm imp}=45$ deg. Masses of the target and the impactor were set of Phobos and Deimos, respectively. The numerical code is the same as that used in \cite{Hyo17c} that was originally developed in \cite{Gen12}. Regarding the EOS, Murchison EOS was used \citep{Nak22}. The total number of SPH particles was $N \simeq 1.1 \times 10^{5}$.

Figure \ref{fig_impact} shows the results of our SPH impact simulations (open squares). The masses of the largest remnant ($M_{\rm lr}$; red points), the second largest remnant ($M_{\rm slr}$; blue points), and the debris ring ($M_{\rm ring} = M_{\rm tot} - M_{\rm lr} - M_{\rm slr}$; black points) are shown. We additionally included the results of independent simulations of $\gamma=0.1$, $v_{\rm imp}=100-300$ m s$^{-1}$, and $\theta_{\rm imp}=45$ deg from \cite{Lei12}, where they performed $N$-body impact simulations of rubble-pile bodies (open circles)\footnote{We note that the exact total mass of \cite{Lei12} is $\sim 40$\% of the total mass of Phobos and Deimos. However, because their impact conditions (i.e., $\gamma=0.10$, $v_{\rm imp}=100-300$ m s$^{-1}$, and $\theta_{\rm imp}=45$ deg) are very similar to ours (i.e., $\gamma \simeq 0.14$, $v_{\rm imp}=100-300$ m s$^{-1}$, and $\theta_{\rm imp}=45$ deg), we used their numerical results with the assumption of $M_{\rm tot}=m_{\rm Pho}+m_{\rm Dei}$ in Fig.~\ref{fig_impact} (i.e., the mass fraction of the largest and the second largest remnants to the total mass).}. Here, the $N$-body approach (open circles) may be more appropriate than the SPH approach (open squares) because Phobos and Deimos are considered to be rubble-pile objects and a prominent impact shock with a phase change would not be produced for $v_{\rm imp}=100-300$ m s$^{-1}$ considered here.

Both simulations $-$ our SPH simulations and the $N$-body simulations of \cite{Lei12} $-$ show that the mass of the largest remnant (red points) decreases with increasing the impact velocity, while the mass of the second-largest remnant (blue points) increases with increasing the impact velocity. The remaining mass, defined as the mass of the debris ring (black points), increases with increasing the impact velocity, indicating more impact debris is produced with increasing the impact velocity.

These results indicate that (1) the impacts, in general, significantly reduce the masses of the moons (i.e., indicated by the points below the dashed lines, where the dashed lines represent their original masses), (2) $v_{\rm imp}=100$ m s$^{-1}$ leads to a catastrophic disruption of Deimos (the mass is reduced more than one order of magnitude; see the blue points and the blue dashed line), and (3) $v_{\rm imp}=300$ m s$^{-1}$ significantly reduces the mass of Phobos (nearly one order of magnitude) in addition to that of Deimos, indicating that most of the mass is distributed as a debris ring (black points). 

Typical impacts of $v_{\rm imp}=100-300$ m s$^{-1}$ with $\theta_{\rm imp}=45$ deg, therefore, are not in agreement with the view of \cite{Bag21} $-$ two moons comparable to Phobos and Deimos that are split from an ancestral single moon would tidally evolve to the orbital configurations of Phobos and Deimos we see today $-$ and imply that the evolution after the hypothetical splitting is not as simple as it was envisioned. Subsequent gravitational and collisional interactions between partially disrupted (and/or catastrophically disrupted) moons and particles in the debris rings, although it is beyond the scope of this study, need to be carefully considered.

Changing the impact angle, $\theta_{\rm imp}$, changes the degree of disruption. However, either Phobos (target here) or Deimos (impactor here) would be significantly disrupted, forming a debris ring, for $v_{\rm imp}=100-300$ m s$^{-1}$. This is because here $v_{\rm imp} \gtrsim 10v_{\rm esc}$ \citep[e.g., see the dependence on the impact angle in][]{Lei12}. For example, if the impact is grazing, it could significantly disrupt the impactor (smaller one), while the target (larger one) could be less disrupted compared to the case of a $45$-deg impact. Thus, changing the impact angle would not change the above conclusion -- Both Phobos and Deimos cannot be intact after the high-velocity impact.

Assuming a progenitor is a rubble-pile object, the particle size distribution of the impact debris may not significantly change from that of the original constituent particles, although it is not directly extracted from the impact simulations. This is because impacts with $v_{\rm imp} \simeq 100-300$ m s$^{-1}$ would not cause noticeable melting and vaporization of the impacted materials. Only around the impact point, particles may be damaged and fragmentation may occur.

Lastly, we note that our SPH simulations and $N$-body simulations of \cite{Lei12} neglect, e.g., the material strength and frictions. Including these additional effects may quantitatively change the masses of the impact remnants, especially for small bodies as small as a few kilometers and less \citep[e.g.,][]{Ben99,Jut10}. However, it is expected that the disruptive outcomes (here $v_{\rm imp} \gtrsim 10v_{\rm esc}$) and the dependence on the impact velocity $-$, i.e., a higher impact velocity results in a more disruptive outcome \citep[e.g.,][]{Lei12} $-$ do not qualitatively change, validating our conclusion above.

\begin{figure*}[t!]
	\plotone{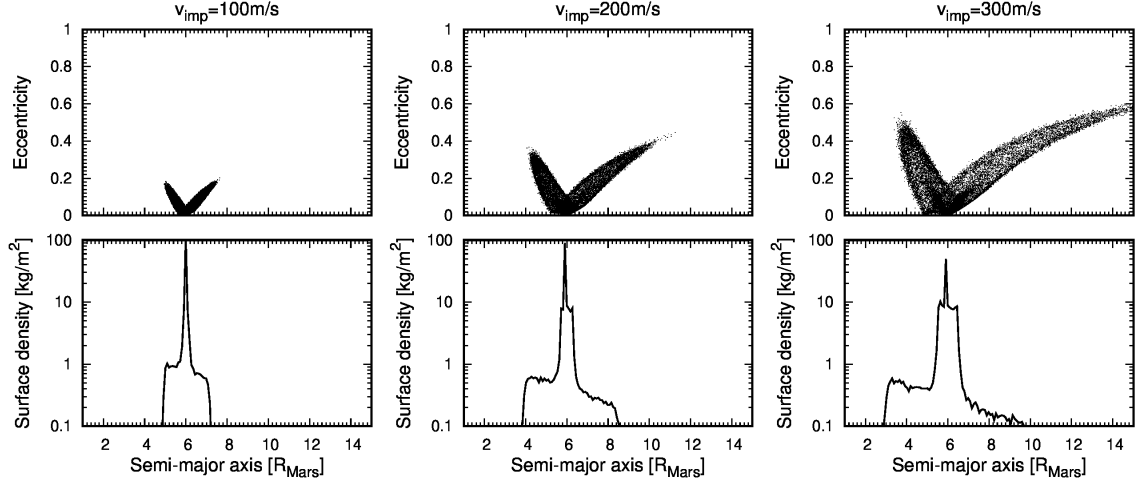}
\caption{Orbital elements, $a$ and $e$, of the debris particles obtained by using the data of SPH simulations (top panels) and corresponding surface densities using the equivalent circular orbital radius, $a_{\rm eq}$, (bottom panels). From left to right panels, cases of $v_{\rm imp}=100$, $200$, and $300$ m s$^{-1}$ are shown. \label{fig_debris}}
\end{figure*}

\section{Discussion}\label{sec_discussion}

\subsection{The successive evolution of the remnant fragments and the debris ring} \label{sec_successive}

As demonstrated in Sec.~\ref{sec_impact}, a disruptive impact between the two moons that are split from a single moon occurs, forming a few large fragments and a debris ring. Using the data obtained from the 3D SPH simulations in free space (i.e., positions and velocities of the debris particles), we constructed the orbits of the debris particles around Mars \citep[a similar approach was used in the Moon-forming giant impact of][]{Jac12}. The top panels of Figure \ref{fig_debris} show the orbits of the debris particles around Mars ($a$ and $e$) for cases of $v_{\rm imp}=100$, $200$, and $300$ m s$^{-1}$. To produce the figure, we assumed that the center of the masses of the two colliding moons orbits around Mars at the Martian synchronous radius ($a_{\rm sync} = 6R_{\rm Mars}$) with eccentricity $e=0$. We assumed that the impact happens in the Martian equatorial plane (i.e., $z=0$ and thus particles have $i \sim 0$), followed by the assumption of \cite{Bag21} that putative Phobos and Deimos formed near the Martian equatorial plane. For the statistical arguments, the debris particles were isotopically distributed in the impact plane ($xy$-plane) to take into account the nature of the isotropic impact direction in the $xy$-plane \citep{Jac12,Hyo18b}. Figure \ref{fig_debris} indicates that most of the debris is concentrated around the synchronous orbit, suggesting that the debris indeed forms a ring-like structure.

Such debris particles would experience a successive dynamical evolution through collisions and gravitational interactions among particles. During the inelastic collisional evolution, the eccentricities are damped, while the angular moment of particles is conserved. The equivalent circular orbital radius, $a_{\rm eq}$, defined as the circular orbit while conserving the angular momentum of a Keplerian orbit with an initial non-zero eccentricity, is given as
\begin{equation}
	a_{\rm eq} = a_{\rm ini} \left( 1 - e_{\rm ini}^2 \right) ,
\label{eq_aeq}
\end{equation}
where $a_{\rm ini}$ and $e_{\rm ini}$ are the initial semi-major axis and eccentricity, respectively. 

Now, using $a_{\rm eq}$, we may estimate the surface density of the debris when the eccentricities are damped to zero. The bottom panels of Fig.~\ref{fig_debris} show surface densities using the data obtained from the SPH simulations. Most of the mass is concentrated within $\sim 5-7R_{\rm Mars}$. Peaks seen at around $\sim 6R_{\rm Mars}$ indicate the largest remnant, which depends on our chosen size of the bins\footnote{Here, we used equally spaced $100$ bins between $1-10R_{\rm Mars}$. Thus, the peak of the surface density becomes, for example, $\sim M_{\rm lr}/(2 \pi a_{\rm syn} \Delta a) \sim 100$ kg m$^{-2}$, where we used $M_{\rm lr}=5 \times 10^{15}$ kg, $a_{\rm syn}=6R_{\rm Mars}$, and $\Delta a = 9R_{\rm Mars}/100$.}. A small number of particles are further distributed in a wide range of the radial direction ($\sim 3-10R_{\rm Mars}$). The arguments presented here, using $a_{\rm eq}$, is an extreme case where the collisional damping is most efficient. In reality, accretion would also take place while collisional damping occurs. To do so, a full $N$-body simulation is required to understand the fate of the debris ring, which is beyond the scope of this paper. 

The key message from Fig~\ref{fig_debris} is that the debris ring would be distributed with a radial width of $ \gtrsim 1R_{\rm Mars}$. The total mass of the debris is only the sum of those of Phobos and Deimos, indicating that the Hill sphere of the total debris mass around Mars ($\sim 38$ km at $a_{\rm syn}=6R_{\rm Mars}$, assuming it is a single object) is about two orders of magnitude smaller than the radial width of the debris ring or less ($\Delta a_{\rm ring} > R_{\rm Mars} \simeq 3390$ km). From this simple consideration, it is expected that more than three moons would accrete from the debris ring because the radial separation of bodies reaching the isolation mass is $\sim 5-10$ times the Hill sphere \citep[e.g.,][]{Kok95}. This separation is still an order of magnitude smaller than the ring width. 

Furthermore, moons accreted in a ring tend to have small eccentricities and the tidal evolution is not efficient especially outside the Martian synchronous orbit, likely leaving the system of multiple moons in the same configuration as it was formed over billions of years. Such an outcome differs from the Martian moon system we see today where only Deimos exits beyond the Martian synchronous orbit. 

This is the reason why the formation of a large ancient inner moon accreted from an inner debris disk $-$ produced within the Martian Roche limit presumably by a giant impact \citep{Hyo17a,Hyo17b} $-$ was proposed for the formation of Phobos and Deimos, i.e., the mean motion resonances of a large single inner moon swept up an outer debris disk concentrated around the Martian synchronous radius, forming only two moons -- Phobos and Deimos -- at specific radial locations \citep{Ros16}. Alternatively, \cite{Can18} considered a less massive extended disk formed by a small impactor compared to that in \cite{Ros16}. This disk spawned transient multiple small inner moons (still massive compared to Phobos and Deimos) that rapidly tidally decayed and did not perturb Phobos and Deimos who naturally accreted from the outer regions of the disk.

Therefore, it seems challenging that only Phobos and Deimos accrete from a debris ring without any external influence (e.g., resonances and/or tides). Instead, multiple small moons would form, i.e., a completely different Martian moons system from the one we observe today.

\subsection{Tidal evolution of the moons} \label{sec_tides}
In this study, we ignored the tidal evolution of the moons that changes their semi-major axes, eccentricities, and inclinations. The tidal evolution of inclination over billions of years is not prominent, while the changes in the pericenter and apocenter distances (a function of the semi-major axis and eccentricity) are not negligible \citep[][their panel (a) in Figure 1]{Bag21}.

A crude estimate, then, can be made for the rate of changes in the apocenter and pericenter distances of Phobos and Deimos, respectively ($\dot{a}_{\rm apo,Pho}$ an $\dot{a}_{\rm per, Dei}$, respectively) as $\dot{a}_{\rm apo,Pho} \sim 4R_{\rm Mars}/10^{9} \sim 1.4 \times 10^{-2}$ m year$^{-1}$ and  $\dot{a}_{\rm per,Dei} \sim 1R_{\rm Mars}/10^{9} \sim 3.4 \times 10^{-3}$ m year$^{-1}$ for Phobos and Deimos, respectively \citep[see][]{Bag21}.

When the tidal evolution is significant enough so that the radial difference in the apocenter distance of Phobos and the pericenter distance of Deimos becomes comparable to the size of the larger of the two moons (in this case, $r_{\rm Pho} \simeq 11.3$ km of Phobos), the orbits of the two moons no longer cross. This occurs with the timescale longer than $\sim r_{\rm Pho}/(\dot{a}_{\rm apo,Pho} + \dot{a}_{\rm peri,Dei}) \sim 6.5 \times 10^{5}$ years. Therefore, in this study, we neglected the effects of tides in our orbital integrations of $<10^{4}$ years.

\subsection{Other challenges in \cite{Bag21} scenario}

In this study, we showed that two moons split from a hypothetical progenitor quickly re-collide and are disrupted into much smaller moons (Sec.~\ref{sec_results} and Sec.~\ref{sec_impact}; see also Fig.~\ref{fig_impact}). One may wonder if the progenitor could be a larger object and the two moons were also larger than \cite{Bag21} considered. Correspondingly, the largest two impact fragments (i.e., in Fig.~\ref{fig_impact}) from the disruptive collision could become Phobos and Deimos, although a complex interplay between the large fragments and small debris needs to be carefully studied (see Sec.~\ref{sec_successive}). However, if this is the case, it already completely changes the picture that \cite{Bag21} envisioned.

More importantly, the physical process of the putative splitting of the progenitor envisioned in \cite{Bag21}, in the first place, seems unlikely. For only two large fragments to be formed (here as Phobos and Deimos), their putative initial ejection velocities (at the time of splitting, i.e., just after the impact) should be comparable to their mutual escape velocity \citep[e.g.,][]{Ben99}. A higher ejection velocity indicates that the impact was more energetic and a larger number of smaller fragments were formed, and vice versa.

\cite{Bag21}, however, envisioned that only two impact fragments existed (as Phobos and Deimos) at the same time their putative ejection velocities (a few hundred meters per second; see Eq.~(\ref{eq_ini}) and Fig.~\ref{fig_initial}) were much larger than their mutual escape velocity (about ten meters per second). Thus, from the above consideration, this situation seems physically unlikely.

Furthermore, \cite{Bag21} envisioned that the two moons orbit near the Martian equatorial plane. This implicitly assumed that the putative impact and the splitting occurred near the Martian equatorial plane. However, the nature of the impactor to the progenitor should be isotropic. From the statistical consideration, the probability that the orbit of the colliding object lies close to the equatorial plane is low.

Although each of the above processes may need to be studied in detail, a number of challenges in \cite{Bag21} scenario already exist. Together with our results -- putative two split moons (as Phobos and Deimos) initially on equatorial, eccentric, and crossing orbits would likely quickly collide --, we conclude that \cite{Bag21} scenario is unlikely.\\

\section{Summary}\label{sec_summary}
\cite{Bag21} envisioned that Phobos and Deimos directly originate from a splitting of a single ancestral moon at around the Martian synchronous orbit ($\sim 6R_{\rm Mars}$) a few billion years ago. At the time of splitting, Phobos and Deimos were envisioned to have moderate eccentricities and orbit near the Martian equatorial plane. Their semi-major axes were assumed to be located inside and outside the synchronous orbit, respectively, followed by a tidal evolution that led to the orbital configuration we see today.

By performing orbital integrations of Phobos and Deimos that are hypothetically formed by the splitting, we found that the two moons likely collide each other during the successive $<10^{4}$ years, and a collision results in a disruptive outcome, forming a debris ring at around the Martian synchronous radius. This process occurs much faster than the tidal forces can evolve moons' orbits away from intersection. The width of the debris ring is $\gtrsim R_{\rm Mars}$ and thus multiple small moons are likely to accrete. This evolutionary path differs from that envisioned by \cite{Bag21} and would form a different moons' system from the one we observe today. Therefore, we conclude that Phobos and Deimos are unlikely to split directly from a single ancestral moon. 

In the coming 2024, Martian Moons eXploration (MMX), developed by the Japan Aerospace Exploration Agency (JAXA), is expected to be launched. The MMX mission plans to collect a sample of $>10$ g from Phobos's surface and return to Earth in 2029 with the aims of elucidating the origin of Martian moons \citep{Fuj19,Usu20}, collecting geochemical information about the evolution of Martian surface environment \citep{Hyo19}, and searching for traces of Martian life \citep{Hyo21}. Therefore, theoretical studies including ours will be finally tested by the MMX mission.\\

\noindent R.H. acknowledges the financial support of MEXT/JSPS KAKENHI (Grant Number JP22K14091). R.H. also acknowledges JAXA's International Top Young program. H.G. acknowledges the financial support of MEXT/JSPS KAKENHI (Grant Number 21H04514, 20KK0080).


\bibliography{Phobos}{}
\bibliographystyle{aasjournal}



\end{document}